\documentstyle[preprint,prb,aps]{revtex}
\begin{document}
\draft
\title{Anomalies in the antiferromagnetic phase of metamagnets}
\author{M. Pleimling and W. Selke}
\address{Institut f\"ur Theoretische Physik B, Technische Hochschule,
D--52056 Aachen, Germany}
 
\maketitle
 
\begin{abstract}
Motivated by recent experiments on the metamagnet FeBr$_2$, anomalies
of the magnetization and the specific heat in the antiferromagnetic 
phase of related spin models are studied systematically using
Monte Carlo simulations. In particular, the dependence of the
anomalous behavior on competing intralayer interactions, the
spin value and the Ising--like anisotropy of the 
Hamiltonian is investigated.
Results are compared to experimental findings on FeBr$_2$.\\ 

\end{abstract}


\noindent {\bf I. Introduction}
 
FeCl$_2$ and FeBr$_2$ are much studied metamagnets of Ising type.
\cite{bir,fert,kin,stry} The
magnetic field ($H$)--temperature ($T$) phase diagram
displays an antiferromagnetically ordered phase, with the transition
to the paramagnetic phase being of first order at low temperatures
and of second order at higher temperatures and lower fields. In the
antiferromagnetic phase the spins of the iron ions are aligned
ferromagnetically in the triangular layers perpendicular to the
c--axis; along that axis there is an antiparallel ordering of the
spins.

In FeCl$_2$, the two kinds of transition meet at a tricritical
point. For FeBr$_2$, a possible decomposition of the tricritical
point into a critical end point and a bicritical end point has
been discussed, in the context of the recent experimental
discovery of lines of anomalies in the antiferromagnetic phase.
\cite{klee1,kat1,kat2,klee2,sel1,hevo,sel2} In
particular, the specific heat as well as the
temperature derivative of the magnetization and the order parameter
may display, at fixed field and varying temperature, shoulders or maxima
below the transition to the paramagnetic phase.

The anomalies have been attributed \cite{sel1,sel2} to two 
crucial ingredients of FeBr$_2$, the effectively weak ferromagnetic
intralayer couplings, due to competing antiferromagnetic 
longer--range interactions, and the highly coordinated interlayer
couplings to many equivalent iron ions in adjacent layers, due
to the superexchange mediated by the non--magnetic bromide planes.
The anomalies have been suggested to reflect the onset of local
fluctuations of a second antiferromagnetic phase, the AII phase,
which, if becoming eventually thermally stable, would
lead to a decomposition of the tricritical point.

In this article, we shall extend the previous analyses to study
quantitatively the dependence of the anomalies on different
parameters of a realistic model \cite{yel,pou} for FeBr$_2$, namely
the competing intralayer couplings, the spin value (being 1 in
FeBr$_2$), and on the Ising--like anisotropies in the Hamiltonian.
Furthermore, the relation between the decomposition of the tricritical
point and the anomalies will be discussed, in particular when comparing
our Monte Carlo results to recent experimental data and their
interpretation.

The layout of the paper is as follows: The Hamiltonian, obtained
from spin wave measurements, is introduced and experimental
findings are outlined in Sect. II. Then results on related Ising
models are presented, clarifying the influence of the competing
interactions and the spin value, followed by a section on the
anisotropic Heisenberg model. In Sect. V, the comparison to
experiments is given. Finally, a brief summary concludes the
article.\\

\noindent{\bf II. Realistic Hamiltonian for FeBr$_2$}
 
The compound FeBr$_2$ has the hexagonal structure shown
in Fig.\ 1, with the magnetic iron ions forming triangular
layers perpendicular to the c--axis (corresponding to the 
z--axis of Cartesian coordinates). Based on spin--wave analyses 
\cite{yel,pou}, the low temperature magnetic properties
of FeBr$_2$ may be obtained from an effective anisotropic
Heisenberg Hamiltonian for the iron ions

\begin{equation}
{\cal H} = \sum\limits_{i > j} \left\{ - \frac{1}{\eta} J_{ij} S_i^z S_j^z
- J_{ij} \left( S_i^x S_j^x + S_i^y S_j^y \right) \right\}
+ \sum\limits_i D \left\{ \left( S_i^z \right)^2 - \frac{2}{3} \right\} - H \sum\limits_i S_i^z
\end{equation}
with the spin value $S=1$. The first term describes exchange interactions
between spins in the same triangular layer and adjacent layers.
Two different sets of interactions have been proposed for the
intralayer couplings, with ferromagnetic nearest neighbor interactions, $J_1$,
and competing antiferromagnetic interactions, extending either up to only
next-nearest neighbors \cite{yel}, $J_2$,

\begin{equation}
J_1/k_B = 7.3 ~\mbox{K} ~~\mbox{and} ~~J_2/k_B = -2.4 ~\mbox{K}  
\end{equation}
or up to third neighbors in the triangular layers \cite{pou}, $J_3$,

\begin{equation}
J_1/k_B = 4.8 ~\mbox{K}, ~J_2/k_B = -0.1 ~\mbox{K} ~~\mbox{and} ~~J_3/k_B 
= -1.0 ~\mbox{K}.
\end{equation}

The interlayer coupling has been determined unambiguously to be

\begin{displaymath}
J'_t/k_B = -2.9 ~\mbox{K}
\end{displaymath}
denoting the total exchange to the adjacent iron layer. Taking into
account the ten equivalent superexchange paths, as mediated by the
bromide planes, each individual bond between neighboring layers
is expected to contribute $J'/k_B = -0.29$ K (see Fig.\ 2).

The Ising--type anisotropy, $\eta = 0.78$ \cite{yel}, in the
first term of the Hamiltonian (1) is enhanced by the second term, 
describing a single-ion anisotropy with the easy axis of the spin along
the z--axis. Here, $D$ is the energy difference between the doublet and
singlet in the lowest triplet of an iron ion, with
$D/k_B = -10.7$ K (for the intralayer couplings of Eq. (2)) or
$D/k_B = -12$ K (for the intralayer couplings of Eq. (3)).

The third term in Eq. (1) describes the effect of the magnetic field,
$H$, applied along the c--axis, i.e. in z--direction.

Fig.\ 3 shows the $H$--$T$ phase diagram of FeBr$_2$ determined from
measurements of the magnetization \cite{klee1,kat2}, dynamic
susceptibility \cite{klee1}, and specific heat. \cite{kat1}
Varying temperature, at fixed field, all three
quantities or their temperature derivatives display in the
antiferromagnetic phase unusual behavior in the form of shoulders
or maxima at about the same temperature $T_a(H)$,
locating the anomaly line. That line seems to evolve from the
tricritical point. Note that it has been alternately suggested 
\cite{kat2} that the anomaly line represents, at sufficiently
large magnetic fields, a true phase
boundary line between different antiferromagnetic orderings. In
that case, one may expect, from mean--field considerations, the
anomaly line to emerge from the (bi)critical point at the end of the
additional phase boundary line \cite{sel2}, with the
tricritical point having turned into a critical end point.

In Fig.\ 3, $T_p$ denotes a line in the paramagnetic phase at which
the dynamic susceptibility \cite{klee1} and the specific heat 
\cite{kat1} show a maximum, when changing temperature at fixed
field. It may seem to be conceivable that this line also evolves
from the tricritical point (or critical end point), but this aspect
has not been investigated experimentally in detail.

In the following, we shall study simplified models based on the 
anisotropic Heisenberg Hamiltonian for FeBr$_2$, Eq. (1), to 
clarify which of its features may enhance (or weaken)
the anomalies and, possibly, decompose the tricritical
point. So far, previous recent analyses \cite{sel1,sel2,her} dealt
with Ising variants of (1), where $S=1/2$.  Perhaps most importantly,
the crucial importance of the high interlayer coordination,
driving the system close to a mean--field type behavior and thereby
inducing local thermal excitations
of AII--type for weak intralayer exchange couplings, was
established.\cite{sel1,sel2} Taking merely interactions to
the geometric nearest neigbor spins in adjacent layers, no
anomalies were found \cite{sel1,sel2} (in that case, fluctuations also destroy
the AII phase, and hence the tricritical point does not 
decompose, as had been seen in simulations \cite{herlan}).
Here, we shall elaborate systematically on
the role of the other parameters in the Hamiltonian, specifically on
that of the competing intralayer exchange couplings, the spin
value, and the spin anisotropy. Thereby, we shall approach
a rather realistic description of FeBr$_2$. A full analysis
of the complete model, Eq. (1), is, however, beyond the
scope of our study. In addition, such an analysis
may provide only an integral and thence a rather limited insight into the
relevant ingredients leading to anomalies in the antiferromagnetic
phase of metamagnets.\\
 
\noindent{\bf III. Ising models}
 
We shall first approximate the Hamiltonian (1) by
Ising models, to elucidate quantitatively the importance of
the competing intralayer couplings as well as
the spin value in stabilizing the anomalies
in the magnetization and the specific heat.\\

{\bf A. Spin 1/2}

Let us consider the  $S=1/2$--Ising Hamiltonian

\begin{equation}
{\cal H} = - J' \sum\limits_{\left< NN \right>} S_i S_j - J_1 \sum\limits_{NN} S
_i S_j
- J_2 \sum\limits_{NNN} S_i S_j - J_3 \sum\limits_{3NN} S_i S_j - H \sum\limits_
i S_i
\end{equation}
where $S_i$ is an Ising spin on site $i$, with spin value 1/2. The
exchange interactions $J_{i,j}$ describe, as before, intralayer (extending
up to third neighbors in the triangular planes, $J_1$, $J_2$, and
$J_3$) and interlayer (to the ten equivalent sites in the adjacent
plane, $J'$) couplings. The couplings are normalized by setting
$|J'| =1$. 
To study the effect of the competing interactions
in the planes, we usually fix the nearest neighbor interaction
$J_1$, and vary the two remaining
antiferromagnetic couplings, $J_2$ and $J_3$. According to the
two different types of exchange constants determined experimentally
\cite{yel,pou}, two cases are of special interest: (a) $J_2 = 0$,
$J_3 < 0$, and (b) $J_2 < 0$, $J_3 = 0$, respectively. To quantify 
the efficiency of the antiferromagnetic couplings in weakening the
effective ferromagnetic nearest neighbor interactions, we also
investigated the case (c) $J_2 = 0$ , $J_3= 0$, changing $J_1$.\\

We simulate systems with $K$ layers, each one consisting
of $L \times L$ spins, using full periodic boundary conditions.
Typically, we choose $K = L = 20$ (to check finite size effects,
$K$ and $L$ ranged from 10 to 40). For equilibration, 10$^4$ Monte
Carlo steps per site (MCS) were used; averages were taken over the
following $2 \times 10^4$ MCS. To improve the statistics and 
to calculate error bars, we performed simulations for ten
realizations, with different random numbers, at a given field, $H$,
and temperature, $T/|J'|$. We computed several quantities of interest,
in particular the energy, $E$, the specific heat, $C$ (both from
energy fluctuations and by differentiating the energy with respect to
the temperature), the magnetization per layer, $M(i)$, and related
quantities such as the total magnetization, $M$, the sublattice
magnetizations, $M_1$ and $M_2$, referring to the odd and even
layers, and the order parameter $M_s = (M_1 - M_2)/2$. To 
take into account phase shifts or flips of entire spin layers,
we usually computed the absolute values of the total magnetization
and the order parameter (which will be denoted by $M_s$ in
the following).
In a few cases, we also determined correlation lengths from 
standard spin--spin correlation functions.\\ 
 
In case (a), the ground state, at $T = 0$
and $H < H_{c0} = 20 \left| J' \right|$,
is the antiferromagnetic structure, $M_1 = 1$ and $M_2 = -1$,
assuming $|J_3| < \frac{1}{2} J_1$ (otherwise, more complicated spin
configurations are stabilized \cite{tan}, due to the competing
interactions along the axes of the triangular layers). Results
of the simulations for that case, fixing the field at $H = 0.9 ~ H_{c0}$
and changing the temperature, are depicted in Fig.\ 4, showing the
specific heat, the order parameter, and the temperature derivative
of the total magnetization for various values of $J_3$.
In accordance with the experimental findings \cite{pou} for
FeBr$_2$, $J_1$ has been set equal to $ 16.75 ~|J'|$ (recall that
the values obtained from the spin wave analysis are $J_1/k_B = 4.8/\eta = 6.2$ K
and $J'/k_B = -0.29/\eta = -0.37$ K).\\

In the finite Monte Carlo system, the transition
to the paramagnetic phase, at $T_N$, manifests
itself, for instance, by a maximum in the specific heat and
a drastic decrease in the order parameter $M_s$, leading to
singularities in the thermodynamic limit. More interestingly, 
anomalous behavior is seen in Fig.\ 4 to occur well below that
transition. For example, the specific heat and the temperature
derivative of the magnetization display shoulders or maxima, becoming
more pronounced with increasing antiferromagmetic interactions $J_3$.
The anomalies vanish for smaller values of $J_3$ (not shown in
Fig.\ 4).\\

Let us briefly recall the physical picture underlying the anomalies,
as has been obtained from mean--field theory of Ising metamagnets
with only nearest neighbor ferromagnetic intralayer couplings. \cite{sel2}
If those couplings are sufficiently weak, compared to the interlayer
interactions, a second antiferromagnetic
phase, AII, may be formed in between the usual antiferromagnetic
phase (AI, with $M_1 > 0$ and $M_2<0$) and the paramagnetic
phase ($M_1 = M_2$), in which both sublattice magnetizations
are positive, but different. The AII phase may
be thought of balancing the conflicting
tendencies of the external field and the antiferromagnetic
interlayer couplings, by maintaining, rather small, clusters
of 'minus' spins in the even layers. Strong ferromagnetic
intralayer interactions tend to disfavor those clusters, thereby suppressing
the AII phase. The transition between the AI and AII phases
is of first order, with the boundary line evolving
from the critical end point on the border line to the paramagnetic
phase, and terminating at a (bi)critical point. From that
(bi)critical point, a line of anomalies emerges. However, such a
line may persist even when there is
a tricritical point, provided the ferromagnetic intralayer couplings
are still sufficiently small.\\

Including now competing antiferromagnetic intralayer couplings,
one may try to cast them, together with $J_1$, in an effective
nearest neighbor ferromagnetic interaction, $J_{eff}$. 
To elucidate the effect of $J_3$ on reducing
$J_{eff}$, we compared our simulational data, case (a), to those
for models with only nearest--neighbor ferromagnetic intraplane
couplings, case (c), varying $J_1$, 
with $|J'| = 1$. In particular, we
determined the change in $J_1$, $\delta J_1$, needed to
reproduce the N\'{e}el temperature $T_N$, at $H = 0$, when $J_3 \ne 0$.
A naive argument of mean--field type suggests that 
$\delta J_1 = J_3$, i.e. $J_{eff} = J_1 + J_3$. In reality,
the antiferromagnetic coupling is much more efficient in 
lowering the effective interaction (as  
may be already seen from the analysis of
the ground states). For instance, at $J_1 = 16.75 ~ |J'|$ and
$J_3 =4.9 ~J'$, we
find $\delta J_1 \approx - 8.15 ~ |J'|$. Note that an even much
stronger reduction in $J_1$ is required in reproducing, instead
of $T_N$, 
the kind of anomalies present when the
antiferromagnetic intraplane couplings are included, see Fig.\ 5.
We find, at $H = 0.9 ~ H_{c0}$, that the value of
$J_3 =4.9~ J'$ then corresponds to weakening $J_1$ from 16.75 $|J'|$ to
roughly  $1.5~ \left| J' \right|$. The high efficiency of $J_3$
(or $J_2$, see below)
in lowering the effective ferromagnetic coupling $J_{eff}$ and hence 
the ferromagnetic ordering in the layers
is crucial, together with the large interlayer coordination,
in explaining the experimentally
found anomalies in FeBr$_2$.\\

Note that the anomalies shown in Fig.\ 4 do not
correspond to sharp phase transitions. For instance, they 
do not seem to give rise to singularities, as one increases the
size of the Monte Carlo systems (going from $L = K = 10$ to
40, e.g., the height of the 'anomalous' maximum in C
below $T_N$, at $H = 0.95~ H_{c0}$,
does not change significantly, in contrast to the behavior of $C$ close
to $T_N$, where the peak becomes clearly visible at $L = K = 20$,
increasing furthermore for the larger systems), nor
is there any indication of hysteresis (by crossing
the anomalies from different directions in the field--temperature
phase plane).
Indeed, the anomalies may be interpreted as reflecting
the onset of local ordering of AII--type, with the long--range
order of the AII--phase being, possibly always, destroyed by fluctuations.
They may be also illustrated by
monitoring typical equilibrium Monte Carlo
configurations.\\

Fixing $J_3$ and varying the external field, one may map the
anomaly line, $T_a(H)$. Examples for a specific value of $J_3$,
$J_3/J_1 = - 0.29$, chosen 
to be close, but, in order to identify easily the location of 
the anomalies,  somewhat larger than that 
obtained for FeBr$_2$,
are depicted in Fig.\ 6. Obviously, the anomalies become stronger
upon increasing the field $H$. However, they seem to go over into
singularities only at the tricritical point on the phase boundary
to the paramagnetic phase, as concluded from analyses of the
types mentioned
above, for fields in the range in between $0.8 ~ H_{c0}$ and $0.95 ~ H_{c0}$
(strictly speaking, if there
is a decomposition of the tricritical point, then the
critical end point on the transition line to the paramagnetic phase and
the critical point at the end of the phase boundary between the 
AI and AII phases would be very close to each other). The
Monte Carlo data for locating the anomaly line are summarized 
in Fig.\ 7, depicting the phase diagram in
the $H$--$T$ plane, at $J_3/J_1 = - 0.29$. $T_a$ has been determined from
the anomaly in the specific heat, in good 
agreement with the corresponding estimates
obtained from the magnetizations.\\  

We also identified the line, $T_p$, in the paramagnetic phase, at which
the specific heat $C$, at fixed fields, displays a maximum as a function
of temperature, see Fig.\ 7. The maximum is believed to reflect a
disordering in the triangular layers. \cite{her2} In close agreement with the
experimental findings on FeBr$_2$, see Fig.\ 3, the line seems to evolve
from the tricritical point. This feature may be, however, accidental. In
mean--field theory, $T_p$ intersects the boundary of the antiferromagnetic
phase, $T_N$, at some point, which is, in general, not related to the
tricritical or critical end point. In the simulations, the height of the
maximum in $C$ does not change drastically on approach to the boundary
of the antiferromagnetic phase, indicating a non-critical behavior,
see Fig.\ 8. It may be wortwhile to clarify this aspect by determining the
location of $T_p$ for different values of $J_3$, where it may be easier
to disentangle the intersection point, of $T_p$ and $T_N$, and the
tricritical point. Note that the specific heat $C$, fixing the temperature and
varying the field, exhibits in the paramagnetic phase
a maximum at about $T_p$ as well, see Fig. \ 7,
in accordance with recent experimental
findings. \cite{klee3}\\   

In case (b), i.e. $J_2 < 0$, $J_3 = 0$, similar conclusions hold.
To describe experimental data on FeBr$_2$, we may choose
$J_1 = 25.1~ |J'|$ and $J_2 = 8.4~ J'$. \cite{yel} The 
antiferromagnetic couplings lead to a weakening of an effective
ferromagnetic intralayer interaction, giving eventually rise to AII--type 
excitations in the even or 'minus' layers which cause the anomalies in the
specific heat $C$ and magnetizations. Actually, $J_3$ is slightly more
efficient than $J_2$ in reducing $J_{eff}$, as seen when
adjusting $J_2$, with $J_1 = 25.1~ |J'|$, to reproduce
$T_N(H=0, J_1 =16.75 ~|J'|, J_3=4.9 ~J')$. The N\'{e}el temperature
is realized, when $J_2 = 10.3 ~ J'$ (being not far from the experimentally
determined value). The anomalies for the two sets
of parameters do not differ much, see Fig.\ 5, demonstrating that
both types of couplings, $J_2$ as well as $J_3$, have a comparable
effect on the anomalies, although they are of quite distinct
physical character (frustration on triangles, $J_2$, or
competition along the axes of the triangular planes, $J_3$). It should
be emphasized that, in general, frustration or competition is not really needed
for obtaining the anomalies in the antiferromagnetic phase of metamagnets:
$J_{eff}$ has to be sufficiently weak.\\ 

In addition, we determined in which way the ratio of the tricritical
temperature $T_t$ to the N\'{e}el temperature $T_N(H=0)$ depends on the
strength of the antiferromagnetic intraplane interactions, for the
cases (a) and (b), fixing
$J_1$ at the value appropriate for FeBr$_2$. The ratio decreases
with increasing $J_2$ or $J_3$ (i.e. decreasing
$J_{eff}$, see also results from mean--field theory, simulations
and high--temperature series expansions \cite{kin,sel2,lan,stan}).
For instance, in case (a), the ratio varies in between about 0.6 and
0.4, when changing $J_3$ from $3.3~ J'$ to $6.5 ~ J'$, with $J_1 = 16.75 ~ |J'|$.
Similarly, the ratio may be lowered to about 0.32, when increasing $J_2$
to $12.1~ J'$, with $J_1 = 25.1 ~ |J'|$.\\

{\bf B. Spin 1}

We now consider the $S=1$ Ising Hamiltonian, see Eq. (4), where each spin
can take the values 0, 1 or --1. Compared to the situation with 
$S=1/2$, thermal fluctuations are facilitated, reducing the transition
temperatures and resulting in more pronounced anomalies.\\

In particular, we studied the case $J_1= 16.75 ~ |J'|$ and $J_3 = 4.9 ~ J'$,
with $J_2 = D = 0$, setting $|J'| =1$, as before for $S=1/2$. Results are
displayed in Fig.\ 9, showing the specific heat $C$
versus temperature at various fields, compare
to Fig.\ 6. Clearly, at larger fields a maximum shows up below $T_N$;
that anomalous behavior in $C$ is corraborated
by similar properties of the magnetizations,
for instance, of $dM_s/dT$.\\

The intersection point of the line of anomalies and the boundary to the
paramagnetic phase is supposedly the tricritical point (again, we found
no evidence for a transition of first order between the AI and AII phases).
The ratio $T_t/T_N(H=0)$ is roughly 0.5, as is the case for $S=1/2$ with
the same values of $J_1$ and $J_3$. Note that the N\'{e}el temperature, $T_N(H=0)$,
is, however, compared to its value for the $S=1/2$ Ising model, lower by
nearly 30 percents. Expressing the coupling constants in terms of 
Kelvin, one easily sees that one moves in the case of $S = 1$ much closer
towards the experimentally determined N\'{e}el temperature in FeBr$_2$,
see below.\\
 
\noindent{\bf IV. Anisotropic Heisenberg models}

We now proceed to the anisotropic $S=1$ Heisenberg model, given in
Eq. (1). In a semiclassical description \cite{viga} of such 
a model, the z--component of the spin, of lenght 1, is discretized,
taking the values $S^z$ = 0, 1 or --1. If $S^z = 0$, then the spin
can rotate, like a classical vector, in the xy--plane, see Fig.\ 10.\\

As in the Ising case, we studied especially the case $J_1 = 16.75 ~ |J'|$,
$J_2 = 0$, and $J_3 = 4.9 ~ J'$. Putting $D= 0$ and $\eta =0.78$, the
simulational data for the specific heat and the magnetizations are
very close to those for the corresponding $S = 1$ Ising model. For
example, at $H= 0.9 ~ H_{c0}$, the critical and anomaly temperatures
are, in the Heisenberg model, lower by roughly one percent. The specific
heat is essentially identical to that shown in Fig.\ 9
for the Ising model. In turn, the
derivative of the order parameter, $dM_s/dT$, for the Heisenberg model,
see Fig.\ 11, agrees very well with that for the Ising case.\\

By turning on the single ion anisotropy, $D$, the critical temperature
is shifted towards higher values, and the anomalies are somewhat
suppressed. In effect, by discriminating $S^z = 0$, one approaches
the $S=1/2$ Ising model.\\   
    
In general, the thermal properties of semiclassical $S=1$ Heisenberg
models seem to resemble quite closely those of the corresponding
$S=1$ Ising models. Deviations are due to spins with vanishing
z--component, which may provide, e.g., additional energy contributions. 
Obviously, the different spin components are not decoupled, leading, perhaps, 
to intriguing effects. However, it is beyond the
scope of our study, to explore this class of models 
extensively (in passing, we may mention our simulational results
on the $S=1$ Heisenberg model with ferromagnetic couplings between
neighboring spins on a square lattice. They indicate non--critical
energy contributions stemming from the xy--components of the spins,
leading to a minor lowering in the transition temperature, compared
to that of the corresponding Ising model).\\ 

Note that the discretization of the z--component of the spin is crucial
in reproducing the anomalous behavior found in FeBr$_2$. A classical
Heisenberg model with spins of fixed length, but arbitrary orientation,
is not expected to show any tendency towards forming, even locally,
the AII phase. Indeed, preliminary simulations on such Heisenberg
models did not show anomalies in the specific heat or the 
magnetizations.\\

\noindent {\bf V. Comparison with experiments}
  
A typical phase diagram of a simplified, but supposedly 
rather realistic model for FeBr$_2$ is depicted
in Fig.\ 7. Obviously, it resembles quite closely the experimental phase
diagram, see Fig.\ 3. However, for a quantitative comparison, a few points
need to be viewed with care.\\

Experimentally \cite{fert,klee1,kat1}, the N\'{e}el
temperature $T_N(H=0)$ is found
to be 14.2 K. Using the two sets of coupling parameters as obtained from
spin wave analyses, see Eqs. (2) and (3),
$T_N(H=0)$ moves towards that
temperature from above, by going from the Ising models with $S=1/2$
to those with $S=1$ and finally to the anisotropic Heisenberg models (both
sets give only slightly different transition temperatures).
Indeed, the $S=1/2$ Ising models, for both sets of parameters, overestimate
$T_N(H=0)$ by almost a factor of 2 (note that previous analyses for FeBr$_2$
were restricted to that case). For the semiclassical S=1 Heisenberg
model, $T_N(H=0)$ is
about 20 K, i.e. it is still too high. That remaining
difference may be partly due to a temperature dependence in the effective
strength of the single--ion anisotropy, $D$, as had been observed in
FeCl$_2$ \cite{bir}, with $D$ becoming smaller at higher temperatures, thereby
tending to lower the N\'{e}el temperature
(in FeBr$_2$, $D$ had been estimated only at a
single, low temperature).--
Similarly, the ratio of $T_t/T_N$ is not reproduced quantitatively by
the model description. While it is about 0.34 in FeBr$_2$, the simulations
yield such low values, e.g., when increasing the antiferromagnetic
intraplane interactions beyond the experimentally determined values, as
discussed above.\\

Of course, the deviations from the experimental results might be due to
simplifications in the model and its treatment,
such as neglect of dipolar interactions between the spins (their
relevance may be seen from the broad two--phase region at low temperatures;
they also would affect the problem of distinguishing the external, used
in experiments, from
the internal magnetic field, used in the simulations) and neglect of much of
the quantum nature of the spins.\\

As stated before, the main aim of our study is
to discuss the origin and character
of the anomalies in the antiferromagnetic phase. While the model description
gives no evidence for a sharp transition from the AI to the AII phase,
such a possibility has been suggested recently based on measurements
of the specific heat \cite{kat1} and, using neutron scattering techniques, the
order parameter $M_s$ \cite{kat2}.
In particular, the specific heat, as a function of temperature, showed
a sharp peak superposed on the broad shoulder or maximum well below the 
transition to the paramagnetic phase \cite{kat1}, becoming sharper with 
increasing field. However, these findings have been questioned
later \cite{klee3}. Indeed, no peaks were detected, but only the
shoulders or maxima, in agreement with the model calculations.\\

In addition, the experimental data for $M_s$ \cite{kat2}
were interpreted in favour of a real transition between
the AI and AII phases. The data, at different fields, were 
plotted against $T/T_N(H)$ and against $T/T_a(H)$. \cite{kat2} In
the former case, data separation was observed for $T < 0.95 ~ T_N$, while in 
the latter case, the data seemed to fall on one 'universal'
curve for $T < T_a$.\\

In Fig.\ 12, we show the corresponding plots of the Monte Carlo data
for the $S=1/2$ Ising model with $J_1 = 16.75 ~ |J'|$, $J_2 = 0$ and
$J_3 =4.9 ~ J'$. Indeed, the behavior is quite similar to that
found in the experiments, with a clear separation of
the order parameter for different fields in the
predicted ranges of temperatures.
Because our analyses, see above, give no
indication for a sharp phase transition at $T_a$, at least for
the fields shown in Fig.\ 12, we, however, tend to conclude that this type
of data presentation is not suitable for providing convincing evidence for
the suggested phase transition.\\

\noindent {\bf VI. Summary}
 
Motivated by recent experiments on the metamagnet FeBr$_2$, anomalies
in the antiferromagnetic phase of Ising--type models, closely
related to the realistic Hamiltonian for that magnet as determined
from spin wave analyses, have been
studied using Monte Carlo techniques.\\

We clarified which ingredients of the Hamiltonian are
relevant for the anomalous properties, such as broad shoulders
or maxima in the specific heat and magnetizations well below the
transition to the paramagnetic phase. In general, the
anomalies can be attributed to local
thermal excitations of the AII phase, due to high coordination of
spins in adjacent layers and  
weak effective ferromagnetic intraplane couplings.
We demonstrated quantitatively, that, in an Ising description,
the anomalies are enhanced by competing
antiferromagnetic intralayer interactions, extending
up to third neighbors, and by the spin value, $S=1$ in the case 
of FeBr$_2$. Going from the Ising
model to an anisotropic $S=1$ semiclassical Heisenberg model leads to
only minor changes in the specific heat and the magnetizations
parallel to the direction of the applied field. The discretization
of the z--component of the spin plays a major role in obtaining
the anomalies.\\

Finally, we compared results of our simulations to experimental 
findings. The simulational data suggest that the anomalies usually do
not correspond to a sharp phase transition. Conflicting 
interpretations of experiments may be viewed with much care.\\
 
\noindent {\bf Acknowledgements}
 
It is a pleasure to thank W. Kleemann, Ch. Binek and O. Petracic
for very useful discussions on their stimulating experiments
on FeBr$_2$. We should like to thank H. Aruga Katori for sending
us intriguing experimental results prior to publication, and 
M. M\"onnigmann for calculations supplementary to the previous
analysis of the mean--field theory.\\

\begin{figure}
\caption{Sketch of the crystal structure of FeBr$_2$, showing the
Fe$^{2+}$ (full circles) and Br$^-$ (open circles) ions.}
\label{fig1} \end{figure}
 
\begin{figure}
\caption{The triangular iron planes, with the ten equivalent neighbors
(full symbols) in the adjacent layer below.}
\label{fig2}
\end{figure}
 
\begin{figure}
\caption{Approximate experimental phase diagram, based on measurements
of the specific heat and magnetizations, see Refs. 5 and 6.
$T_a$ denotes the anomaly line in the antiferromagnetic phase,
$T_p$ indicates the location of maxima in the specific heat, at fixed
magnetic fields, in the paramagnetic phase, and $T_N$ the boundary
to the paramagnetic phase. At low temperatures, the transition
of first order leads to a 2-phase region.}
\label{fig3}
\end{figure}
 
\begin{figure}
\caption{Monte Carlo data of (a) the specific heat $C$,
(b) the order parameter $M_s$, and (c) the temperature derivative of the
total magnetization $d|M|/dT$ versus $T/|J'|$ for
the $S=1/2$ Ising model with $J_1 = 16.75 ~ |J'|$ and $H = 0.9 ~ H_{c0}$ at various
values of $J_3$. Systems with $K = L = 20$ spins
are considered. Here and in the following figures, error bars are only
shown when they are larger than the sizes of the symbols.} 
\label{fig4}
\end{figure}
 
\begin{figure}
\caption{Simulational data of the specific heat $C$ versus temperature
$T/|J'|$, at $H =0.9 ~ H_{c0}$, with (a) $J_1 = 16.75 ~ |J'|$, 
$J_3 =4.9 ~ J'$, (b) $J_1 = 25.1 ~ |J'|$, $J_2 = 10.3 ~ J'$, and (c)
$J_1 = J_{eff} = 8.6 ~ |J'|$, where the N\'{e}el temperature $T_N(H=0)$
is approximately the same in all three cases. Systems with $K = L =20$ 
spins are considered.}
\label{fig5}
\end{figure}
 
\begin{figure}
\caption{The specific heat $C$ as a function of temperature $T/|J'|$
for the $S=1/2$ Ising model with $J_1 = 16.75 ~ |J'|$ and $J_3 = 4.9 ~|J'|$
at various fields. Systems with $K = L = 20$ spins
are simulated.}
\label{fig6}
\end{figure}
 
\begin{figure}
\caption{Phase diagram in the field ($H$)-- temperature ($T$)
plane of the $S=1/2$ Ising model with $J_1 =16.75 ~ |J'|$ and $J_3 = 4.9 ~J'$.
The anomalies, $T_a$, in the antiferromagnetic phase, determined from
the specific heat and the magnetizations, are denoted by full circles.
The maxima in the specific heat ,$T_p$, at fixed fields (open) or fixed
temperatures (full), in the paramagnetic phase are shown by triangles.
Monte Carlo systems with $20 \times 20 \times  20$ spins are simulated.}
\label{fig7}
\end{figure}
 
\begin{figure}
\caption{Specific heat $C$ versus temperature, $T/|J'|$, at
various fields in the paramagnetic phase, see Fig.\ 7.}
\label{fig8}
\end{figure}
 
\begin{figure}
\caption{Specific heat $C$ versus temperature $T/|J'|$ for the $S=1$
Ising model with $K = L = 20$ spins, at $J_1 =16.75 ~ |J'|$
and $J_3 = 4.9 ~ J'$, and various fields.}
\label{fig9}
\end{figure}
 
\begin{figure}
\caption{Orientations of the spin used in the semiclassical $S=1$ Heisenberg
model, with discretization of the z--component and continuous symmetry in
the xy--plane.}
\label{fig10}
\end{figure}
 
\begin{figure}
\caption{Temperature derivative of the order parameter,  $dM_s/dT$, versus
temperature $T/|J'|$ for the anisotropic $S=1$ Heisenberg model with
$20 \times 20 \times 20$ spins, for $J_1 = 16.75 ~ |J'|$, $J_3 = 4.9 ~ J'$,
$D = 0$, and $\eta =0.78$, at $H = 0.9 ~ H_{c0}$.}
\label{fig11}
\end{figure}
 
\begin{figure}
\caption{Order parameter, $M_s$, versus
 reduced temperatures, (a)
$T/T_N(H)$, and (b) $T/T_a(H)$, for the $S=1/2$ Ising model with
$J_1 = 16.75 ~ |J'|$, $J_3 = 4.9 ~ J'$, at various fields, showing
ranges of temperatures (see text) where data (almost) collapse or
are widely separated, as in experiments on FeBr$_2$, see Ref. 7. Systems with
$K = L = 20$ spins are simulated.}
\label{fig12}
\end{figure}

\end{document}